\documentclass[10pt,aps,prb,superscriptaddress,nobibnotes,notitlepage,twocolumn]{revtex4-2}

\usepackage{multirow}
\usepackage{siunitx} 
\usepackage{amsmath}
\usepackage{amssymb}
\usepackage{braket}
\usepackage{mathtools}
\usepackage{graphicx}
\usepackage{color}
\usepackage{xcolor}
\usepackage{xspace}
\usepackage{lipsum}
\usepackage{upquote}
\usepackage{csquotes}
\usepackage{float}
\usepackage[colorlinks=true,allcolors=black]{hyperref}

\DeclareUnicodeCharacter{0650}{\unskip}

\usepackage{color}
\hypersetup{
     colorlinks   = true,
     citecolor    = blue
}

\newcommand{\extfig}[1]{Extended Data Fig.~\ref{#1}}

\usepackage[colorlinks=true,allcolors=black]{hyperref}



\makeatletter
\setcitestyle{numbers,super,sort&compress,open={},close={}}
\makeatother

\begin{document}

\title{Continuous correlated states and dual-flatness in a moiré heterostructure}

\author{Mohammed M. Al Ezzi \textsuperscript{1}}
\email{alezzi@seas.harvard.edu}
\author{Na Xin\textsuperscript{2,3}}
\author{Yanmeng Shi\textsuperscript{2,3}}
\author{Shuigang Xu\textsuperscript{2,3}} 
\author{Julien Barrier\textsuperscript{2,3}}
\author{Alexey Berdyugin\textsuperscript{2,3}}
\author{Shubhadeep Bhattacharjee\textsuperscript{2,3}}
\author{Angelika Knothe\textsuperscript{2,3}}
\author{Kenji Watanabe\textsuperscript{5}}
\author{Takashi Taniguchi\textsuperscript{6}}
\author{Vladimir Falko\textsuperscript{2,3,7}}
\author{Giovanni Vignale\textsuperscript{4}}
\author{Andre K. Geim\textsuperscript{2,3}}
\author{Shaffique Adam\textsuperscript{8,9}}
\email[]{shaffique@wustl.edu}
\author{Kostya S. Novoselov \textsuperscript{4}}
\email[]{kostya@nus.edu.sg}
\author{Minsoo Kim\textsuperscript{10}}
\email[]{minsoo@sogang.ac.kr}

\thanks{\\
\textsuperscript{1}John A.~Paulson School of Engineering and Applied Sciences, Harvard University, Cambridge, Massachusetts 02138, USA. 
\textsuperscript{2}Department of Physics and Astronomy, University of Manchester, Manchester, UK.
\textsuperscript{3}National Graphene Institute, University of Manchester, Manchester, UK.
\textsuperscript{4}The Institute for Functional Intelligent Materials, National University of Singapore, Singapore, Singapore.
\textsuperscript{5}Research Center for Electronic and Optical Materials, National Institute for Materials Science, Tsukuba, Japan.
\textsuperscript{6}Research Center for Materials Nanoarchitectonics, National Institute for Materials Science, Tsukuba, Japan.
\textsuperscript{7}Henry Royce Institute for Advanced Materials, Manchester, UK.
\textsuperscript{8}Department of Physics, Washington University in St. Louis, St. Louis, Missouri, USA.
\textsuperscript{9}Department of Materials Science and Engineering, National University of Singapore, Singapore, Singapore.
\textsuperscript{10}Department of Physics, Sogang University, Seoul, South Korea. \\
}

\date{\today} 
\maketitle

\textbf{Many-body effects in condensed matter yield novel quantum states when the electronic density of states is enhanced. A vivid example is flat bands, which suppress kinetic energy and let interactions dominate, when they are filled with an integer number of electrons in moiré systems. Yet flat bands and commensurate fillings are not the only conditions for correlated phenomena. Situations may occur where the band structure develops locally enhanced density of states, leading to strong correlations even at non-integer fillings, although such cases often yield pseudogaps that make detection elusive. Here we demonstrate that small-angle twisted monolayer–bilayer graphene combines moiré-induced global flat band and additional local band flattening. Their coexistence allows direct comparison of correlated effects. The global route stabilizes commensurate states, while the local mechanism produces nearly flat bands, lifting degeneracy and generating symmetry breaking at non-integer fillings, yet without opening a global gap. Because there is no global gapped signature, the system remains metallic, but the effect reveals itself in anomalous Hall responses, signaling time-reversal symmetry breaking and valley polarization. Our results demonstrate dual-flatness as a guiding principle, extending moiré physics beyond commensurate fillings and identifying topological transport as a probe of gapless correlated metals.}

\begin{figure}[!t]
    \centering
    \includegraphics[width=1.0\columnwidth]{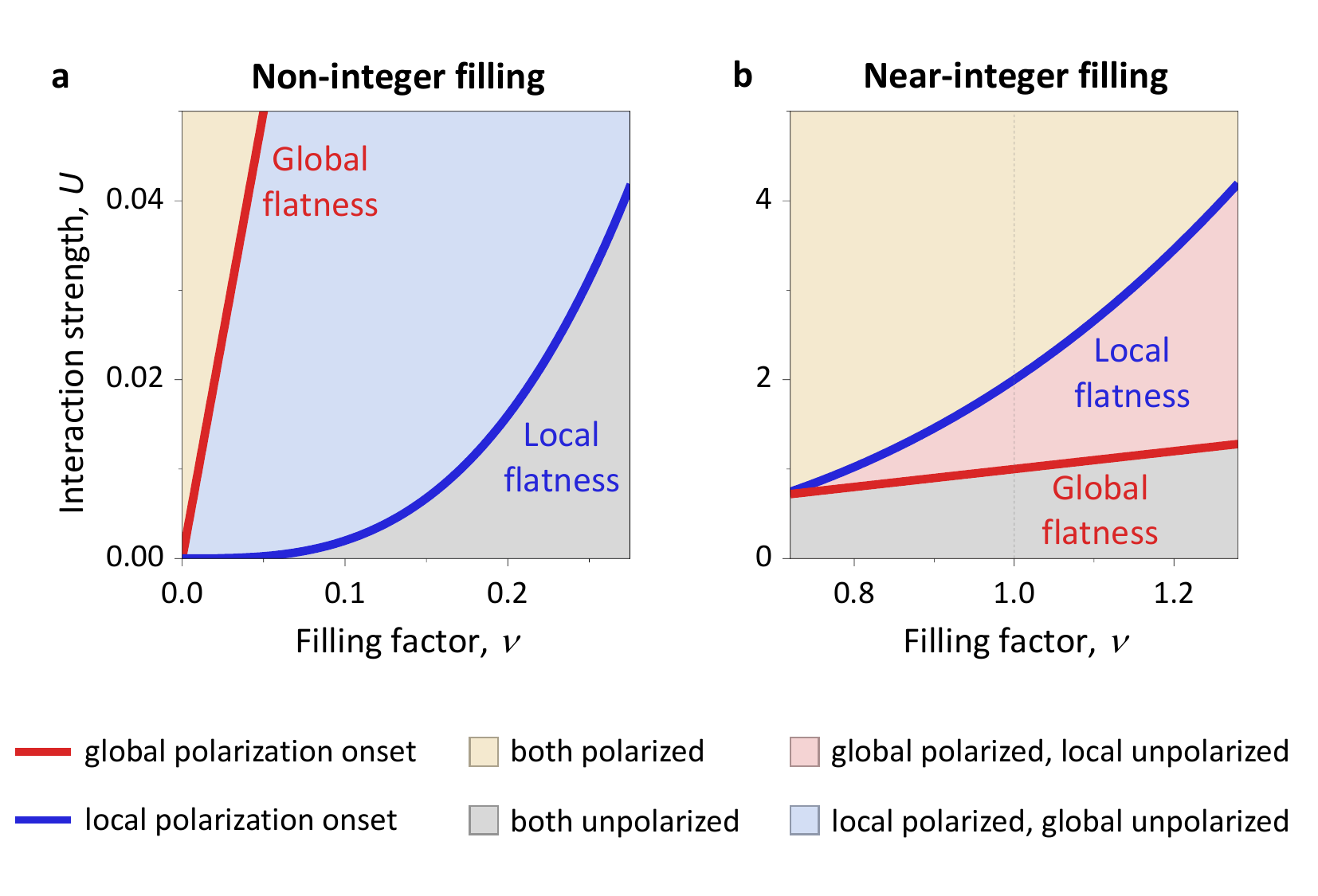}
    \caption{
     \textbf{Analytical Stoner model illustrating the difference between global and local band flatness.} The calculated mean-field phase boundaries reveal a reversal in susceptibility to band polarization depending on filling. In both cases, the band is assumed two-fold degenerate. Away from integer filling (Fig.~\ref{fig0}a), a locally flat band ($E\sim k^4$, bandwidth $W_L \gg U$, blue curve) is more susceptible to band polarization than a globally flat band (bandwidth $W_G \lesssim U$, $E\sim k^2$, red curve) even though its total bandwidth is large. $U$ is expressed in units of $W_G$, with $W_G = 1$. Near integer filling (Fig.~\ref{fig0}b), the situation reverses, and the globally flat band becomes more susceptible. Twisted monolayer--bilayer graphene naturally realizes both types of flatness simultaneously, which accounts for the two distinct classes of correlated states observed at integer and non-integer fillings, respectively.
    }
    \label{fig0}
\end{figure}

\begin{figure*}[!t]
    \centering
    \includegraphics[width=0.95 \textwidth]{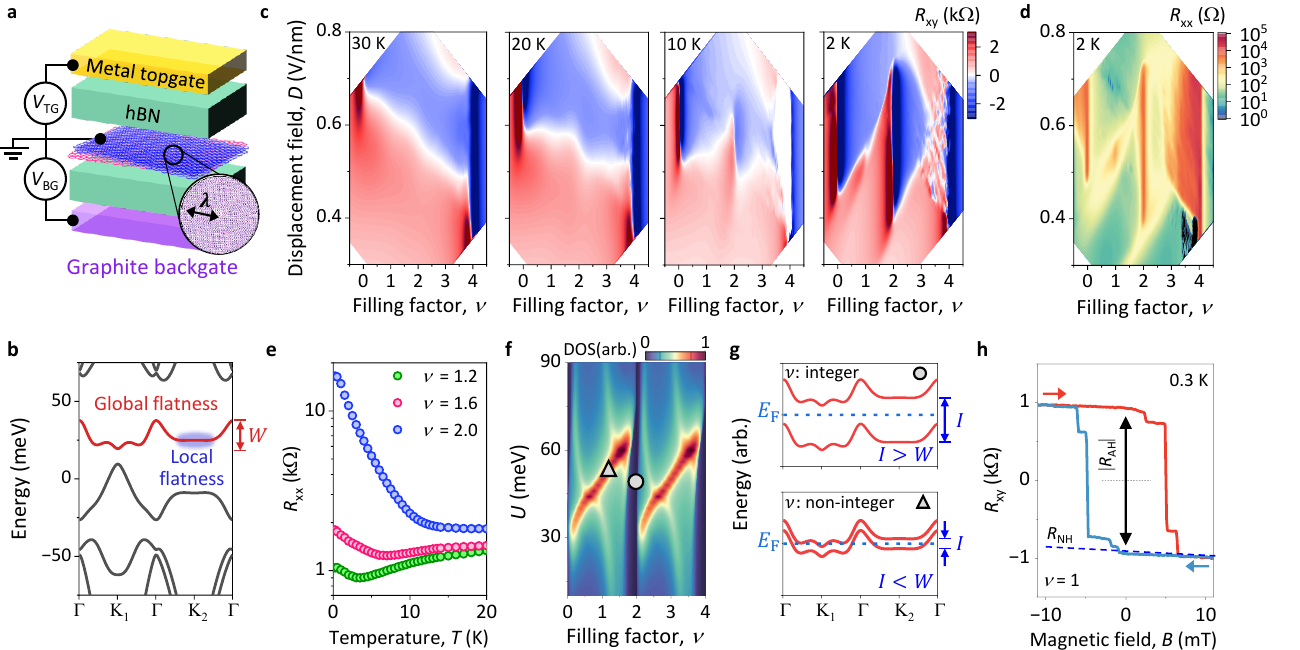}
    \caption{\textbf{Dual-flatness and correlated states in twisted monolayer--bilayer graphene.} 
    \textbf{(a)} Device schematic. Bilayer graphene is stacked on monolayer graphene at a small twist angle and encapsulated in hBN. Carrier density and displacement field $D$ are controlled by a graphite backgate $V_{\mathrm{BG}}$ and a metal topgate $V_{\mathrm{TG}}$. 
    \textbf{(b)} Calculated band structure at finite $D$. A gap isolates the first conduction band (red). The moiré potential compresses the overall bandwidth $W$ (global flatness), while the bilayer subsystem develops a locally flat region near the $K_2$ valley (blue), associated with a van Hove singularity (local flatness). 
    \textbf{(c)} Hall resistance $R_{xy}$ versus $\nu$ and $D$ at 30, 20, 10, and 2~K. At high temperature the neutrality line runs diagonally; upon cooling it reshapes into vertical features at integer fillings $\nu = 1$ and $2$ (\textit{discrete} states) and diagonal features at non-integer fillings (\textit{continuous} states). 
    \textbf{(d)} Longitudinal resistance $R_{xx}$ at 2~K. Enhanced resistance traces both the discrete and continuous correlated features. 
    \textbf{(e)} Temperature dependence of $R_{xx}$ at $\nu = 1.2$, $1.6$, and $2.0$. The $\nu = 2.0$ state is insulating; $\nu = 1.2$ and $1.6$ remain metallic. 
    \textbf{(f)} Calculated density of states in the $\nu$--$U$ plane. Diagonal peaks mark the van Hove singularity in the bilayer subsystem. 
    \textbf{(g)} Schematic band structures at integer filling (top, $\circ$) and non-integer filling (bottom, $\triangle$), corresponding to the marked points in \textbf{(f)}. At $\nu = 2$ the interaction energy $I$ exceeds $W$, opening a full gap. At non-integer fillings $I < W$ and the band splits near $K_2$ without a global gap. 
    \textbf{(h)} Hall resistance at $\nu = 1.0$, decomposed into normal ($R_{\mathrm{NH}}$) and anomalous ($R_{\mathrm{AH}}$) components. The zero-field offset reveals a valley-polarized anomalous Hall effect. Data from a device with twist angle $1.18^\circ$.
}
    \label{fig1}
\end{figure*}

\begin{figure*}[t]
    \centering
    \includegraphics[width=1 \textwidth]{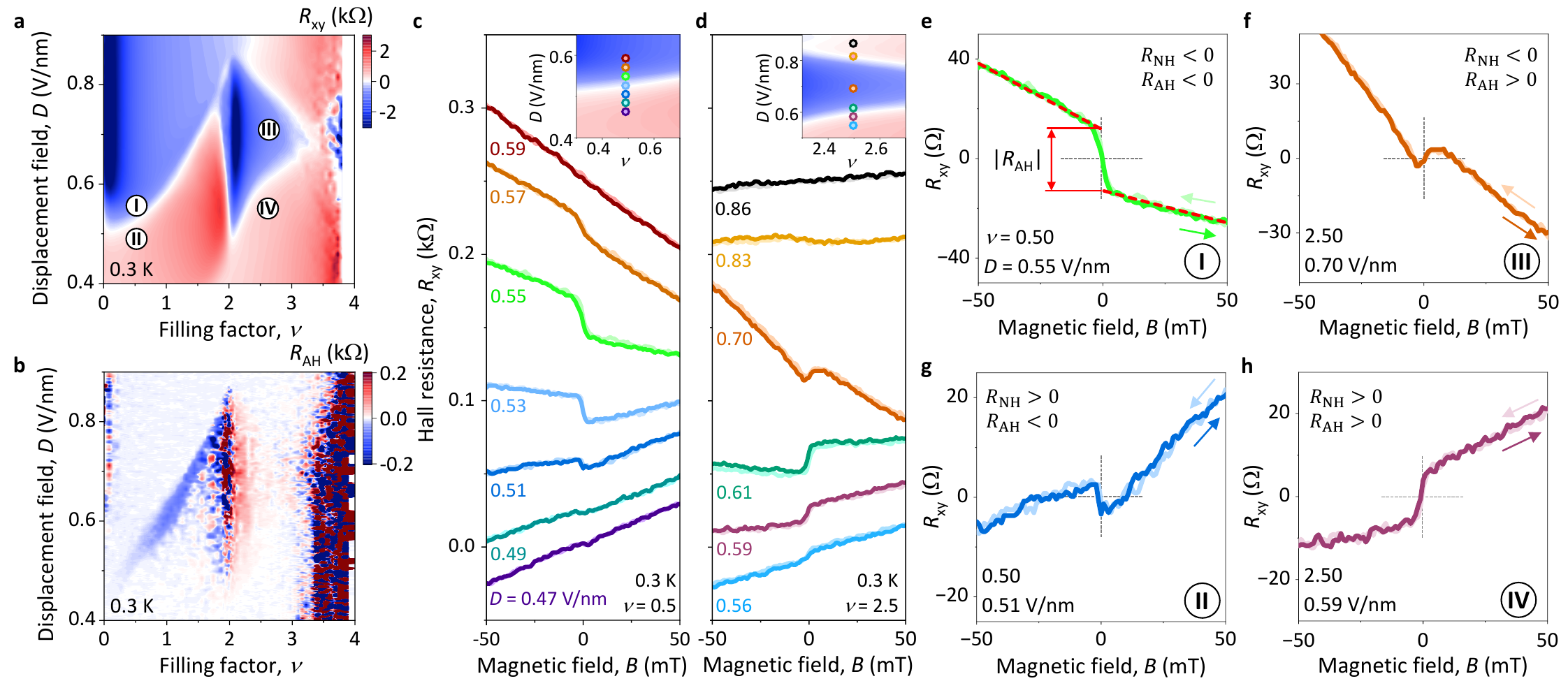}
    \caption{\textbf{Anomalous Hall effect along correlated lines}.
    \textbf{(a)} Hall resistance map at $B = 1$~T and 
    \textbf{(b)} anomalous Hall resistance $R_{\mathrm{AH}}$, both plotted versus $\nu$ and $D$ at 0.3~K. White circles I--IV in \textbf{(a)} mark the positions of the measurements in \textbf{(e--h)}. 
    \textbf{(c,d)} AHE at $\nu = 0.5$ and $\nu = 2.5$ for varying $D$ (color-coded as in the insets) at 0.3~K. Plots are vertically offset for clarity. The AHE peaks near the correlated line and diminishes away from it. 
    \textbf{(e--h)} Hall resistance at four representative points (I--IV), illustrating all four sign combinations of the normal ($R_{\mathrm{NH}}$) and anomalous ($R_{\mathrm{AH}}$) components. $R_{\mathrm{AH}}$ is obtained from the offset between the zero-field intercept and the normal-Hall background (red dashed line). Continuous states ($0 < \nu < 2$) show a consistent $R_{\mathrm{AH}}$ sign; the discrete state at $\nu = 2$ reverses sign. Arrows indicate sweep direction. Data from a device with twist angle $1.30^\circ$.
    }
    \label{fig2}
\end{figure*}

\begin{figure*}[!t]
    \centering
    \includegraphics[width=0.95\textwidth]{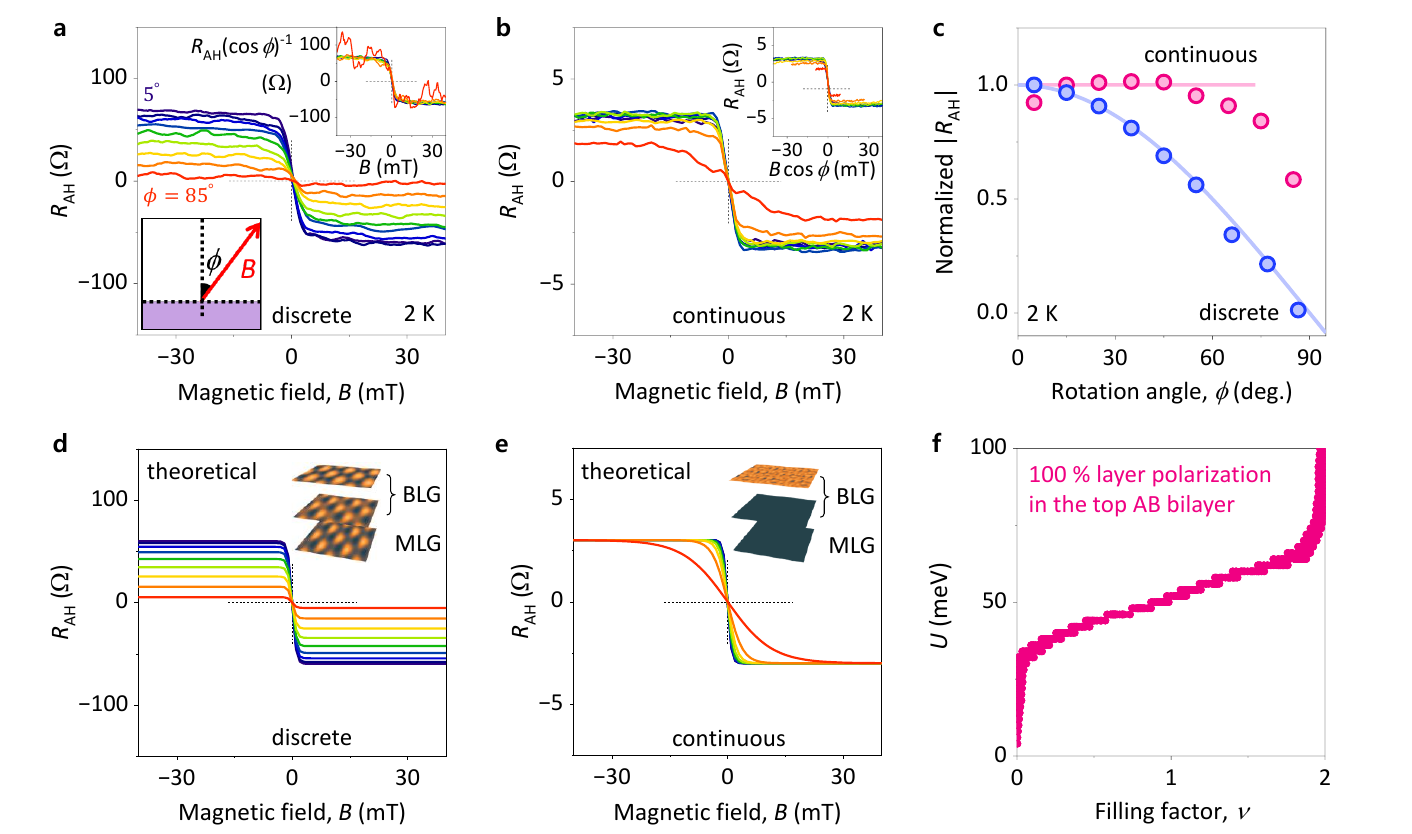}
    \caption{\textbf{Angular dependence of the anomalous Hall effect and its microscopic origin.}
    \textbf{(a,b)} Anomalous Hall resistance measured at 2~K for the discrete correlated state (global flatness) at $\nu \approx 2$ and the continuous correlated state (local flatness) at $\nu \approx 1$, with the field angle $\phi$ varied between $5^\circ$ and $85^\circ$. Main panels show $R_{\mathrm{AH}}$ versus $B$. Insets show the same data rescaled: in \textbf{(a)}, dividing by $\cos\phi$ collapses the traces; in \textbf{(b)}, plotting against $B\cos\phi$ collapses the curves. 
    \textbf{(c)} Normalized $|R_{\mathrm{AH}}|$ (normalized to its value at $\phi = 0^\circ$) versus $\phi$. The discrete state (blue) follows a $\cos\phi$ dependence; the continuous state (red) is nearly angle-independent.
    \textbf{(d,e)} Calculated $R_{\mathrm{AH}}$ curves reproduce the experimental contrast. Insets show layer-resolved wavefunction distributions: delocalized across all graphene layers in the discrete state, but confined to the top layer of the AB bilayer in the continuous state. 
    \textbf{(f)} Calculated layer polarization in the $\nu$--$U$ plane. The trace marks the boundary where polarization in the top layer of the AB bilayer reaches 100\%; other regions have intermediate values.
    }
    \label{fig4}
\end{figure*}

Band flatness relevant to moiré correlated states can arise in two fundamentally different forms. In the first, which we call global flatness, the electronic bandwidth is compressed across the entire Brillouin zone to only a few tens of meV, as commonly realized in moiré superlattices, letting electron--electron interactions dominate and stabilize correlated insulators, superconductors, and fractional quantum anomalous Hall states at commensurate band fillings~\cite{Rafi2011moire, lopes2012continuum, andrei2021marvels, Chen2019evidence, Yu2020independent, Wang2020correlated, Zhao2022moire, Chen2023magnetic, park2023observation, Li2024wigner}. In the second, local flatness, the band retains a large overall bandwidth but develops momentum-confined regions of nearly vanishing dispersion, most prominently near van Hove singularities where the density of states is sharply peaked. Correlation-driven symmetry breaking near such singularities has been observed in Bernal bilayer and rhombohedral multilayer graphene~\cite{zhou2022isospin, de2022cascade, seiler2022quantum, lu2024fractional, Han2025signatures, Patterson2025superconductivity, Choi2025superconductivity, Lu2025extended, Holleis2025fluctuating, pantaleon2023superconductivity}, producing correlated metallic states at non-integer fillings. Although the two mechanisms have been studied separately, a unified framework that places them on equal footing and provides experimental observables to distinguish them has so far been lacking.

To place these two mechanisms on equal footing, we introduce an analytical Stoner model (see Methods) that compares a globally flat band with small total bandwidth $W_G$ and a locally flat band with larger bandwidth $W_L$ but vanishing curvature near the band bottom. The two cases respond differently to filling, as captured by the mean-field phase diagrams shown in Extended Data Fig.~\ref{fig_stoner_methods}. In Fig.~\ref{fig0}, we focus on two contrasting filling regimes. At non-integer fillings (Fig.~\ref{fig0}a), the locally flat band polarizes at parametrically lower interaction strength because its vanishing curvature produces a divergent effective mass that concentrates spectral weight where the electrons reside; at integer fillings (Fig.~\ref{fig0}b), the globally flat band polarizes first because its smaller total bandwidth is easier to overcome. The local case additionally supports a partially polarized metallic phase that has no counterpart in the global case: symmetry is broken, yet the system remains gapless. This filling-dependent competition is the essence of what we call dual-flatness, and it maps directly onto the physics of twisted monolayer--bilayer graphene, where both mechanisms coexist within a single band.

Twisted monolayer--bilayer graphene (tMBG) realizes both mechanisms, typically found in distinct systems, within a single conduction band. The moiré potential compresses the global bandwidth, while the bilayer subsystem introduces a displacement-tunable van Hove singularity that locally enhances the density of states near the $K_2$ valley~\cite{xu2021tunable}. In transport measurements, these two mechanisms manifest as distinct low-temperature features: \textit{discrete} correlated states at integer fillings $\nu = 1$ and $2$, where global flatness opens hard gaps consistent with cascade-like transitions~\cite{zondiner2020cascade, wong2020cascade}, and \textit{continuous} correlated states that extend across non-integer fillings, driven by local flatness and accompanied by spontaneous time-reversal-symmetry breaking probed through the anomalous Hall effect. Tilted-field measurements further reveal that the two types of states possess distinct orbital character, reflecting the different wavefunction geometries imposed by global and local flattening.

We studied tMBG devices with twist angles between $1.1^\circ$ and $1.3^\circ$, fabricated by the cut-and-stack method \cite{Chen2019evidence, Yu2020independent} and encapsulated in hexagonal boron nitride (hBN) (see Methods). Independent tuning of carrier density $n$ and displacement field $D$ was achieved through a graphite backgate and metal topgate (Fig.~\ref{fig1}a); $D > 0$ corresponds to the electric field pointing from the monolayer toward the bilayer. Under a finite $D$, a gap opens between the low-energy bands, isolating the first conduction band (red in Fig.~\ref{fig1}b), which becomes nearly flat over a finite $D$ range \cite{xu2021tunable}. The calculated band structure of this band directly illustrates the dual-flatness regime: the moiré potential compresses the overall bandwidth $W$ to tens of meV, while the bilayer subsystem produces a locally flat region near the $K_2$ valley associated with a van Hove singularity. We measured longitudinal and Hall resistance as functions of filling factor $\nu = 4n/n_0$ (where $n_0$ is the density at full band filling), $D$, and temperature, obtaining consistent results across multiple devices.

Figure~\ref{fig1}c presents Hall resistivity maps as a function of $\nu$ and $D$ at temperatures from 30~K down to 2~K. At 30~K, the electron--hole neutrality line appears as a straight diagonal separating electron- and hole-like regions. Upon cooling, this line reshapes into a zigzag pattern with an overall opposite slope, composed of two distinct types of features: vertical segments at integer fillings $\nu = 1$ and $2$, and diagonal segments extending continuously across non-integer fillings for $0 < \nu < 2$ and $2 < \nu < 4$. Both types emerge within a displacement-field window $0.45 < D < 0.85$~V/nm, forming a set of correlated lines in the $\nu$--$D$ plane that are absent at high temperature.

The temperature dependence of $R_{xx}$ separates the two types sharply (Figs.~\ref{fig1}d,e). At $\nu = 2.0$, $R_{xx}$ shows a pronounced insulating upturn below $\sim\!20$~K, consistent with a gapped correlated phase at integer filling. At $\nu = 1.2$ and $1.6$, $R_{xx}$ decreases with cooling and remains finite down to 2~K, with only a weak low-temperature upturn characteristic of a strongly correlated metal rather than a gapped insulator. The coexistence of insulating discrete states and metallic continuous states within the same band points to two distinct interaction regimes governed by different portions of the band structure.

The microscopic origin of this contrast lies in the displacement-field evolution of the conduction band. As $D$ increases, the moiré superlattice compresses the global bandwidth non-monotonically, reaching a minimum of $\sim\!20$~meV for $0.45 < D < 0.85$~V/nm (\extfig{non-monotonic}), realizing the globally flat limit of the Stoner model (Fig.~\ref{fig0}). Simultaneously, the bilayer subsystem develops a van Hove singularity near the $K_2$ valley (\extfig{bandsEvol}), creating a momentum-localized ultra-flat region where the effective mass diverges (\extfig{valley_polarization}a), the analogue of the locally flat limit. The density of states in the $\nu$--$U$ plane (Fig.~\ref{fig1}f) shows a diagonal peak that maps directly onto the experimentally observed continuous correlated features. When the chemical potential enters this region, exchange interactions lift the fourfold spin-valley degeneracy. At integer fillings, the correlation energy exceeds the $\sim\!20$~meV bandwidth and the resulting band splitting opens a hard gap across the Brillouin zone (Fig.~\ref{fig1}g, top): one band is pulled below the Fermi level at $\nu = 1$, two at $\nu = 2$. Along the continuous features at non-integer fillings, the band splits near $K_2$ but the Fermi level avoids the gap (Fig.~\ref{fig1}g, bottom), producing partial polarization that lowers the exchange energy while preserving metallic conduction. The system thus achieves strong correlations without a global gap, through the locally enhanced density of states alone.

To probe this symmetry breaking directly, we measure the anomalous Hall effect (AHE). The conduction band of tMBG is topological and fourfold degenerate in spin and valley~\cite{xu2021tunable}, carrying Berry curvature that is equal and opposite in the two valleys. When interactions lift the valley degeneracy, this cancellation is broken and a net anomalous Hall signal appears. A representative measurement at $\nu = 1.0$ (Fig.~\ref{fig1}h) illustrates how we separate the Hall response into normal and anomalous components, $R_{xy} = R_{\mathrm{NH}} B + R_{\mathrm{AH}}$. The normal term $R_{\mathrm{NH}}$ gives the $B$-linear background determined by the carrier type and density, while $R_{\mathrm{AH}}$ encodes the integrated Berry curvature of the occupied bands; its sign directly reflects which valley lies lower in energy.

The anomalous Hall response traces the correlated features across the full $\nu$--$D$ phase diagram (Fig.~\ref{fig2}a), where the correlated lines appear as sharp features at which $R_{xy}$ crosses zero. To quantify the effect, we map $|R_{\mathrm{AH}}|$ across the same parameter space (Fig.~\ref{fig2}b). At non-integer fillings ($0 < \nu < 2$), $|R_{\mathrm{AH}}|$ peaks along the continuous correlated lines with a consistent sign, indicating a preferred valley polarization that persists throughout this filling range. At the discrete state $\nu = 2$, the anomalous Hall signal is comparable in magnitude but reverses sign across the line, reflecting a change in valley order at the commensurate gap.

Detailed measurements at $\nu = 0.5$ and $\nu = 2.5$ (Figs.~\ref{fig2}c,d) reveal that the anomalous Hall response is strongly enhanced near the correlated lines and vanishes away from them, confirming that symmetry breaking is confined to the regions of locally enhanced density of states. At four representative points in the phase diagram (I--IV in Fig.~\ref{fig2}a), we resolve all four possible sign combinations of $R_{\mathrm{NH}}$ and $R_{\mathrm{AH}}$, labeled Types~I--IV. At $\nu = 0.5$ (Figs.~\ref{fig2}e,f), tuning $D$ across the correlated line reverses $R_{\mathrm{NH}}$ while $R_{\mathrm{AH}}$ remains unchanged; at $\nu = 2.5$ (Figs.~\ref{fig2}g,h), $R_{\mathrm{AH}}$ takes the opposite sign but is likewise unaffected by crossing the boundary, while $R_{\mathrm{NH}}$ again reverses. The normal and anomalous contributions thus evolve independently, so that carrier type and Berry-curvature sign can each be tuned without affecting the other.

The four sign combinations encode the correlation-induced valley ordering of the fourfold-degenerate conduction band. Consider a horizontal cut at $D = 0.6$~V/nm, varying $\nu$ from 0 to 4 (Fig.~\ref{fig2}a). For $0 < \nu < 2$, the lower subband pair is filled: $R_{\mathrm{NH}}$ indicates electron-like carriers at low filling (Type~I) transitioning to hole-like carriers at higher filling (Type~II), with the boundary at a continuous diagonal line. The anomalous contribution vanishes away from this line but becomes finite in its vicinity, with a negative sign indicating that the K-valley states lie lower in energy than the K$^\prime$-valley states; the first two bands to fill therefore carry K and K$^\prime$ character in that order. The vanishing $R_{\mathrm{AH}}$ away from the line confirms that these two subbands carry opposite valley character. At $\nu = 2$, the upper subband pair begins to fill. For $2 < \nu < 4$, the Hall response repeats the same pattern of electron-to-hole transition (Types~III and IV), but now $R_{\mathrm{AH}}$ exhibits a positive sign near the correlated line, indicating that the K$^\prime$-valley states lie lower; the third and fourth bands thus fill as K$^\prime$ and K. Because K$^\prime$ is the last to fill in the lower pair and the first to fill in the upper pair, the complete sequence is (K, K$^\prime$, K$^\prime$, K). The valley order reverses at $\nu = 2$, where the global gap separates the two subband pairs, whereas within each pair the continuous states preserve valley order.

Our theoretical model captures this valley symmetry breaking quantitatively. The integrated Berry curvature exhibits sharp peaks when the chemical potential crosses the locally flat portion of the band near $K_2$ (\extfig{valley_polarization}). Because of the finite valley $g$-factor, reversing the magnetic field flips the valley ordering and thus the sign of the Berry curvature, producing the field-antisymmetric anomalous Hall response observed throughout the continuous correlated regime (Figs.~\ref{fig2}c,d). This behavior persists down to fields as small as $\pm 10$~mT, confirming that valley polarization is robust and does not require a global gap.

The anomalous Hall effect provides a further experimental handle to distinguish the two types of correlated states. Figures~\ref{fig4}a--c show the angular dependence of $R_{\mathrm{AH}}$ measured at 2~K with the field tilted between $5^\circ$ and $85^\circ$ from the sample normal. For the discrete state near $\nu \approx 2$, where the correlated gap is most robust, the anomalous Hall magnitude decreases systematically with tilt angle and collapses when normalized by $\cos\phi$ (inset, Fig.~\ref{fig4}a), indicating that the magnetization follows the external field. For the continuous state near $\nu \approx 1$, where at 2~K the gap is not well developed and the response is dominated by the locally flat band, the magnitude remains nearly constant across all tilt angles and collapses only when plotted against $B\cos\phi$ (inset, Fig.~\ref{fig4}b), consistent with orbital magnetization pinned out of plane. The contrast is summarized in Fig.~\ref{fig4}c: the discrete state follows a $\cos\phi$ dependence, whereas the continuous state is nearly angle-independent.

The distinct angular responses reflect different orbital confinement geometries. Layer-resolved wavefunction projections (insets, Figs.~\ref{fig4}d,e) show that discrete states are delocalized across all three graphene layers, while continuous states are strongly confined to the top layer of the bilayer subsystem. In the discrete state, this extended distribution allows the orbital magnetic moment to follow the external field, producing the $\cos\phi$ dependence. In the continuous state, confinement to a single layer restricts orbital motion to the out-of-plane direction, locking the magnetization perpendicular to the layers regardless of field orientation. A simple model based on this confinement picture (see Methods) quantitatively reproduces the experimental data (Figs.~\ref{fig4}d,e).

The calculated layer polarization (Fig.~\ref{fig4}f) quantifies this confinement across the $\nu$--$U$ plane. Along the continuous correlated lines, the wavefunction is almost entirely confined to the top layer of the AB bilayer, explaining the fixed out-of-plane magnetization observed experimentally. This strong layer confinement occurs precisely where the density of states peaks (Fig.~\ref{fig1}f) and coincides with the correlated lines mapped in transport (Fig.~\ref{fig2}a), establishing a direct correspondence between local flatness, enhanced interactions, and orbital confinement. In contrast, the discrete states show weight distributed across all graphene layers, consistent with the field-following magnetization.

\emph{Conclusions}---Our work establishes dual-flatness as a design principle for correlated quantum matter, in which moiré-induced global flatness and local flatness coexist within a single band to generate distinct many-body states. This challenges the prevailing view that strong correlations in moiré systems are tied exclusively to commensurate fillings, and establishes that non-integer fillings can host equally rich correlated physics through a distinct mechanism rooted in momentum-space band flattening rather than global bandwidth compression. In twisted monolayer--bilayer graphene, global flatness produces gapped correlated insulators at integer fillings, while local flatness drives a correlated metal at non-integer fillings with robust valley polarization that directly accesses Berry curvature without requiring a global gap. The tilted-field response reveals the microscopic distinction between the two: delocalized wavefunctions at integer fillings yield field-following magnetization, whereas layer-confined wavefunctions at non-integer fillings pin the orbital moment out of plane. Because the two mechanisms are independently tunable through the displacement field, tMBG provides a setting in which bandwidth-driven and curvature-driven correlations can be compared directly. Van Hove singularities that produce local flatness are common in layered moiré systems, suggesting that the interplay between global and local flattening may be relevant beyond the specific system studied here.



\setcitestyle{super,comma,sort&compress}  
\bibliographystyle{naturemag}
\bibliography{Bibliography}

\renewcommand{\figurename}{\textbf{Extended Data Fig.}}
\renewcommand{\thefigure}{\arabic{figure}}
\setcounter{figure}{0}

\clearpage


\noindent \textbf{Methods}\\
\textbf{Device fabrication}\\
Twisted monolayer–bilayer graphene heterostructures were fabricated using a cut-and-stack method. Graphene flakes with monolayer/bilayer steps were mechanically exfoliated onto SiO$_2$/Si substrates. To obtain separated monolayer and bilayer crystals from a single flake, an atomic force microscopy tip was used to cut the original mono–bilayer graphene into two distinct regions. The stacking process was carried out using a polymer stamp consisting of polypropylene carbonate (PPC) supported on a polydimethylsiloxane (PDMS) film. A thin hBN flake was first picked up with the PPC/PDMS stamp, followed by sequential pickup of the bilayer graphene, the rotated monolayer graphene (at a target twist angle of $\sim1.2 \deg$), and finally another hBN crystal. The entire stack was then released onto a thin graphite flake prepared on an oxidized Si substrate, which served as a bottom gate electrode. This resulted in a five-layer heterostructure of the form hBN/monolayer graphene/bilayer graphene/hBN/graphite. To fabricate devices, the heterostructures were patterned into Hall-bar geometries by electron-beam lithography and etched using CHF$_3$/O$_2$ plasma. One-dimensional edge contacts were defined by electron-beam evaporation of Cr/Au (3 nm/60 nm), and additional Cr/Au electrodes were deposited to form the top gate.\\

\noindent \textbf{Electric transport measurements}\\
Electrical transport measurements were performed using a standard low-frequency lock-in technique. The excitation frequency was varied between 5.3 and 30.5 Hz, and the excitation current was kept below 100 nA. For dual-gated devices, the total carrier density $n$ was determined by the combined effect of the top and bottom graphite gate voltages, $n = \frac{1}{e}\left(C_{\mathrm{TG}} V_{\mathrm{TG}} + C_{\mathrm{BG}} V_{\mathrm{BG}}\right)$, where $e$ is the elementary charge, and $C_{\mathrm{TG}}$ and $C_{\mathrm{BG}}$ are the capacitances per unit area of the top and graphite gates, respectively. The capacitances were independently calibrated through magnetotransport measurements. The perpendicular displacement field $D$ was defined as $D = \frac{1}{2\varepsilon_{0}}\left(C_{\mathrm{TG}} V_{\mathrm{TG}} - C_{\mathrm{BG}} V_{\mathrm{BG}}\right)$, with $\varepsilon_{0}$ the vacuum permittivity. A positive displacement field ($D>0$) corresponds to the direction pointing from the monolayer graphene toward the bilayer graphene. The relative twist angle of the monolayer-bilayer graphene was determined from the periodicity of Brown-Zak oscillations, which provided an accurate calibration of the moiré superlattice period.\\

\noindent \textbf{Anomalous Hall effect measurements}\\
The measured Hall resistance is expressed as $R_{xy} = R_{\mathrm{NH}}B + R_{\mathrm{AH}}$, where the normal term $R_{\mathrm{NH}} = \pm (ne)^{-1}$ depends on carrier density $n$ and carrier type, giving the conventional $B$-linear Hall response, and the anomalous contribution $R_{\mathrm{AH}} = \pm (e^2/h)\rho_0^2 \sum_k \Omega_k$ reflects the Berry curvature summed over occupied states in the conduction band. Experimentally, the anomalous part is extracted as a zero-field offset and is strongly enhanced along the correlated lines. A representative trace near $\nu \approx 1$ (Fig.~\ref{fig1}h) shows a zero-field jump in $R_{xy}$ together with magnetic hysteresis, with a coercive field of about $\pm 10$~mT at 0.3~K, indicating a valley-polarized, symmetry-broken ground state stabilized by global flatness. 

To quantify the magnitude, $|R_{\mathrm{AH}}|$ was determined in hysteretic cases from the difference $R_{xy}(B = 0^+) - R_{xy}(B = 0^-)$, where the Hall values are taken at zero field after sweeping $B$ down from positive and negative directions, respectively. For non-hysteretic cases, $|R_{\mathrm{AH}}|$ was obtained from the separation of zero-field intercepts extrapolated from positive and negative field ranges (Fig.~\ref{fig2}e). For constructing Fig.~\ref{fig2}b, $R_{xy}$ was measured at $\pm0.8$~mT, and the difference was taken to estimate $R_{\mathrm{AH}}$, assuming that the normal Hall contribution is negligible at such small fields. This approach yielded clear anomalous signals for $0 < \nu < 2$, but not for $2 < \nu < 4$, where the valley polarization is likely insufficiently reversed within $\pm0.8$~mT. Indeed, Fig.~\ref{fig2}c,d suggest that a stronger anomalous response exists near the correlated line in this region. We therefore attribute the weak signal in Fig.~\ref{fig2}b for $2 < \nu < 4$ to the limited magnetic-field range during the mapping rather than the absence of an intrinsic AHE. These definitions and procedures allow both the sign and magnitude of $R_{\mathrm{AH}}$ to be mapped systematically, providing direct information on the valley order across the $\nu$–$D$ phase diagram.\\

\noindent \textbf{Stoner model for global and local flatness}\\
We present an analytical Stoner model for a one-dimensional two-fold degenerate band that yields exact closed-form phase boundaries contrasting the effects of global and local flatness. While the experimental system we study has a fourfold-degenerate band (spin and valley), the two-fold model retains the essential physics needed to illustrate the key distinction between the two flatness mechanisms. The band dispersion takes the power-law form $\varepsilon(\tilde{k}) = W|\tilde{k}|^p$, where $\tilde{k} = k/\pi \in [-1,1]$ is the normalized momentum and $W$ is the bandwidth. Global flatness is modeled by $p=2$ (parabolic band, bandwidth $W_G$) and local flatness by $p=4$ (quartic band, bandwidth $W_L = rW_G$; Extended Data Fig.~\ref{fig_stoner_methods}a), where the ratio $r = W_L/W_G$ captures the separation of energy scales between the two mechanisms. The local flatness criterion $\partial^2\varepsilon/\partial k^2 \ll Wa^2$, where $a$ is the lattice constant, is satisfied near $k=0$ for the quartic band ($\partial^2\varepsilon/\partial k^2 = 12Wk^2a^4 \ll Wa^2$ when $12(ka)^2 \ll 1$) but never for the parabolic band ($\partial^2\varepsilon/\partial k^2 = 2Wa^2$ everywhere; Extended Data Fig.~\ref{fig_stoner_methods}b).  Following the rigid-band Stoner approach~\cite{ghazaryan2023multilayer}, we write the grand potential as $\Phi/L = E_\text{kin}(n_1) + E_\text{kin}(n_2) + Un_1n_2 - \mu(n_1+n_2)$, where $n_1$ and $n_2$ are the per-flavor fillings and $U$ is the interaction strength. Parametrizing in terms of total filling $\nu = n_1 + n_2$ and polarization $\delta = n_1 - n_2$, the kinetic energy per flavor is $E_\text{kin}(n_f) = Wn_f^{p+1}/(p+1)$. Minimizing the free energy at fixed $\nu$ yields the phase boundaries shown in Extended Data Fig.~\ref{fig_stoner_methods}c,d. For global flatness, a single first-order boundary at $U_c = \nu W_G$ separates the unpolarized from the fully polarized phase. For local flatness, two boundaries emerge: a second-order onset at $U_\text{onset} = (\nu^3/2)W_L$ and a full-polarization boundary at $U_\text{full} = \nu^3 W_L$ (for $\nu \leq 1$), enclosing a partially polarized metallic phase unique to the local case in which the system is polarized yet remains gapless. The phase diagrams are plotted in interaction strength $U$ with $r = 4$ as an illustrative example, so that the physically distinct energy scales of the two mechanisms are directly visible. The qualitative conclusions presented here are robust across a range of values of $r$, with the precise crossover filling at which the local and global phase boundaries interchange depending on the specific choice of $r$.  We note that the power-law dispersions $\varepsilon \propto k^p$ are not intended as realistic models of the full Brillouin zone, but rather as minimal models designed to isolate the essential distinction between curvature-driven correlations at low fillings and bandwidth-driven correlations near commensurate fillings. The comparison between the two band types is therefore most meaningful at small non-integer fillings, where the band-bottom curvature dominates, and near $\nu = 1$, where the overall bandwidth sets the relevant energy scale.\\

\noindent \textbf{Electronic model}\\
\noindent We model twisted monolayer–bilayer graphene (tMBG) within a continuum framework \cite{xu2021tunable,Rafi2011moire}. For the monolayer, considering $p_{z}$ orbitals on sublattices $A_{1}$ and $B_{1}$, the nearest-neighbor tight-binding Hamiltonian is  
\begin{equation}
H_{\text{MLG}} =
\begin{pmatrix}
-\tfrac{U}{2} & \gamma_{0} f(\mathbf{k}) \\
\gamma_{0} f^{*}(\mathbf{k}) & -\tfrac{U}{2}
\end{pmatrix}
\end{equation}
with  $f(\mathbf{k})=\sum_{j=1}^{3} e^{i\mathbf{k}\cdot\delta_{j}},$ where $\delta_{j}$ denote the three nearest-neighbor bond vectors connecting atoms on the $A$ and $B$ sublattices.  

For AB-stacked bilayer graphene, we take $p_{z}$ orbitals on sites $A_{2}, B_{2}, A_{3}, B_{3}$. The Hamiltonian is  
\begin{equation}
H_{\text{BLG}} =
\begin{pmatrix}
0 & \gamma_{0}f(\mathbf{k}) & \gamma_{4}f(\mathbf{k}) & \gamma_{3} f^{*}(\mathbf{k}) \\
\gamma_{0}f^{*}(\mathbf{k}) & \Delta & \gamma_{1} & \gamma_{4} f(\mathbf{k}) \\
\gamma_{4}f^{*}(\mathbf{k}) & \gamma_{1} & U/2+\Delta & \gamma_{0}f(\mathbf{k}) \\
\gamma_{3} f(\mathbf{k}) & \gamma_{4} f^{*}(\mathbf{k}) & \gamma_{0} f^{*}(\mathbf{k}) & U/2
\end{pmatrix}
\end{equation}
with parameters obtained from density-functional calculations \cite{jung2014accurate}:  
\begin{equation}
(\gamma_{0}, \gamma_{1}, \gamma_{3}, \gamma_{4}, \Delta) = (-2.61, 0.361, 0.283, 0.138, 0.015)\,\text{eV}
\end{equation}

The coupling between the monolayer and bilayer is described by three tunneling matrices $T_{j}$ ($j=1,2,3$):  
\begin{equation}
T_{j} =
\begin{pmatrix}
w_{AA} & e^{-i(j-1)\phi}w_{AB} & 0 & 0 \\
e^{i(j-1)\phi}w_{AB} & w_{AA} & 0 & 0
\end{pmatrix}
\end{equation}
where $\phi = 2\pi/3$. The values for the Fourier components are $w_{AA}=0.050$ eV and $w_{AB}=0.085$ eV \cite{adak2020tunable}.  \\

\noindent \textbf{Normal and anomalous Hall resistivity}

\noindent We assume that the components of the Hall conductivity are:
\begin{equation}
\sigma_{xx} = \sigma_{yy} = \frac{\sigma_0}{1 + (\omega_c \tau)^2}
\end{equation}
and
\begin{equation}
\sigma_{xy} = -\sigma_{yx} = \frac{\sigma_0 (\omega_c \tau)}{1 + (\omega_c \tau)^2} + \sigma_{an}
\end{equation}
where $\sigma_0$ is a Drude-like conductivity $\frac{ne^2 \tau}{m}$, $\tau$ is the momentum relaxation time, and $\omega_c = \pm \frac{eB}{m}$ is the cyclotron frequency associated with a magnetic field $B$. Notice that $\pm e$ in this expression is the charge of the carriers – holes or electrons, depending on the position of the Fermi level.

Here the anomalous Hall conductivity is
\begin{equation}
\sigma_{an} = \frac{e^2}{h} \sum_k \Omega_k
\end{equation}
where the sum runs over all the occupied states for a given chemical potential. Inverting the conductivity tensor to the first order in $\omega_c \tau$, we find

\begin{align}
\rho_{xy} &= \frac{\rho_0 (\omega_c \tau) + \rho_0^2 \sigma_{an}}{1 + 2\rho_0 (\omega_c \tau) \sigma_{an} + (\rho_0 \sigma_{an})^2} \nonumber \\
&\simeq \frac{B}{ne} + \frac{e^2}{h} \rho_0^2 \sum_k \Omega_k.
\end{align}

where we have set $\rho_0 = \frac{1}{\sigma_0} = \frac{m}{ne^2 \tau}$. \\ \\

\noindent \textbf{Angular behavior of the correlated states}
\label{appB}

\noindent The anomalous contribution is proportional to the orbital magnetization. We parameterize the magnetization by its saturation value and introduce a functional form to describe its dependence on the magnetic field. This form should be general enough to apply to various cases, such as different angular orientations and different fillings. Therefore, it should depend on the magnetic flux normalized by the magnetization itself, ensuring independence from the strength of the orbital magnetization. One possible form is:

\begin{equation}
R_{\text{AH}} = \frac{h}{e^2} M_{\text{sat}} \mathcal{F}\left[ \frac{\text{flux}}{M_{\text{sat}}} \right]
\end{equation}

Where $M_{\text{sat}}$ is calculated from the Berry curvature, for which we have an expression in the text. One realization for our model is given by:

\begin{equation}
R_{\text{AH}} = \frac{ h}{e^2} M_{\text{sat}} \tanh\left(\frac{B \cos\phi}{M_{\text{sat}}}\right)
\end{equation}

Piecewise Definition: \\

All information is now encoded in the magnetization, and its behavior is governed by the wavefunction localization. The wavefunction of the continuous correlated line case is localized to a single 2D plane, and therefore the total magnetization always points in the out-of-plane direction. On the other hand, the wavefunction of the discrete correlated states is not localized and can point in any direction following the magnetic field. We express the behavior of magnetization for the two different correlated cases as:

\begin{equation}
M_{\text{sat}} =
\begin{cases}
M_{\text{sat}}^{2D} = M_{\text{sat}}^{\text{tot}}, & \text{for continuous case (2D)} \\
\vspace{0.4cm} \\ 
M_{\text{sat}}^{3D} = M_{\text{sat}}^{\text{tot}} \cos\phi, & \text{for discrete case (3D)}
\end{cases}
\end{equation}

After substituting the magnetization in the expression for $R_{AH}$ we get:

Discrete Case:
\begin{equation}
R_{\text{AH}}^{\text{vert}} = \frac{ h}{e^2} M_{\text{sat}}^{tot} \cos\phi \tanh\left(\frac{B \cos\phi}{M_{\text{sat}}^{tot} \cos\phi}\right)
\label{angvert}
\end{equation}

Continuous Case:
\begin{equation}
R_{\text{AH}}^{\text{diag}} = \frac{ h}{e^2} M_{\text{sat}}^{tot} \tanh\left(\frac{B \cos\phi}{M_{\text{sat}}^{tot}}\right)
\label{angdiag}
\end{equation}

Eqs. \ref{angvert} and \ref{angdiag} are used to generate the plots in Fig. \ref{fig4} (d) and (e) respectively. \\

\begin{figure*}[!t]
   \includegraphics[width=0.9 \textwidth]{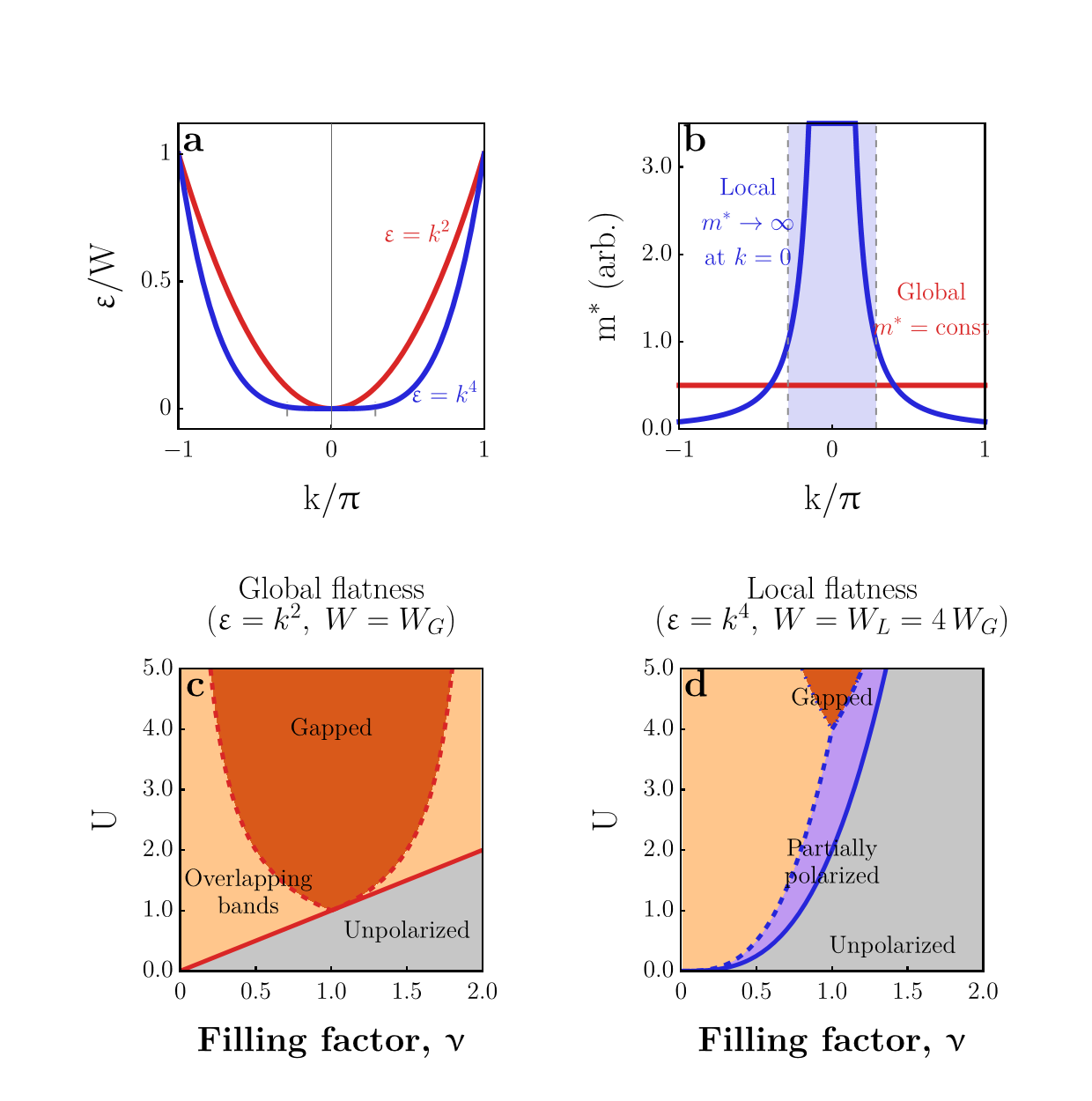}
   \caption{\textbf{Stoner model for global and local flatness: band dispersions and phase diagrams.}
   \textbf{(a)} Analytical 1D band dispersions $\varepsilon(k)/W$ normalized by the respective bandwidth. The $k^2$ band (red) has uniform curvature, while the $k^4$ band (blue) has vanishing curvature near $k=0$ (shaded region, $|k/\pi| \leq 1/\sqrt{12} \approx 0.29$, marking the boundary where $\partial^2\varepsilon/\partial k^2 = Wa^2$).
   \textbf{(b)} Effective mass $m^*(k) \propto [\partial^2\varepsilon/\partial k^2]^{-1}$, encoding the local curvature of each band. The $k^2$ band (red) has constant $m^*$, while the $k^4$ band (blue) has $m^* \to \infty$ at $k = 0$ (shaded region, same as panel (a)), the microscopic origin of correlated behavior at non-integer fillings.
   \textbf{(c,d)} Analytically derived phase diagrams in interaction strength $U$ for global ($W = W_G$, $k^2$) and local ($W = W_L = rW_G$, $k^4$, $r = 4$) flatness, identifying unpolarized (gray), overlapping bands (orange), and gapped (red) regions. The local case additionally hosts a partially polarized metallic phase (purple) with no counterpart in the global case.
   }
   \label{fig_stoner_methods}
\end{figure*}

\begin{figure*}[!t]
   \includegraphics[width=0.6 \textwidth]{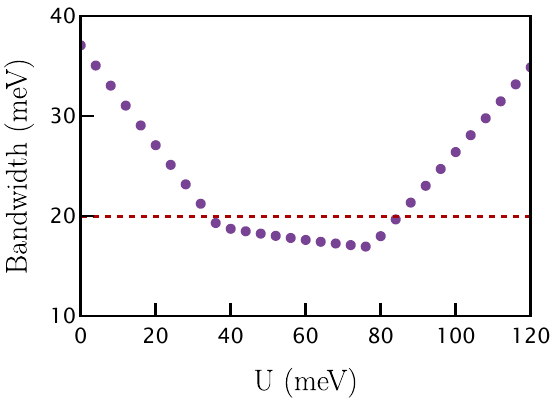}
   \caption{\textbf{Region of Critical Bandwidth}. Bandwidth of the lowest electron miniband as a function of electric potential energy difference (U), exhibiting non-monotonic behavior with a region $(30<U<85)$ meV of minimal bandwidth. The horizontal dashed red line indicates a proposed border between correlated and non-correlated regions, serving as a guide rather than a strict delineation. The calculations are shown for twist angle 1.22$^\circ$. 
  }
   \label{non-monotonic}
\end{figure*}

\begin{figure*}[!t]
   \includegraphics[width=1 \textwidth]{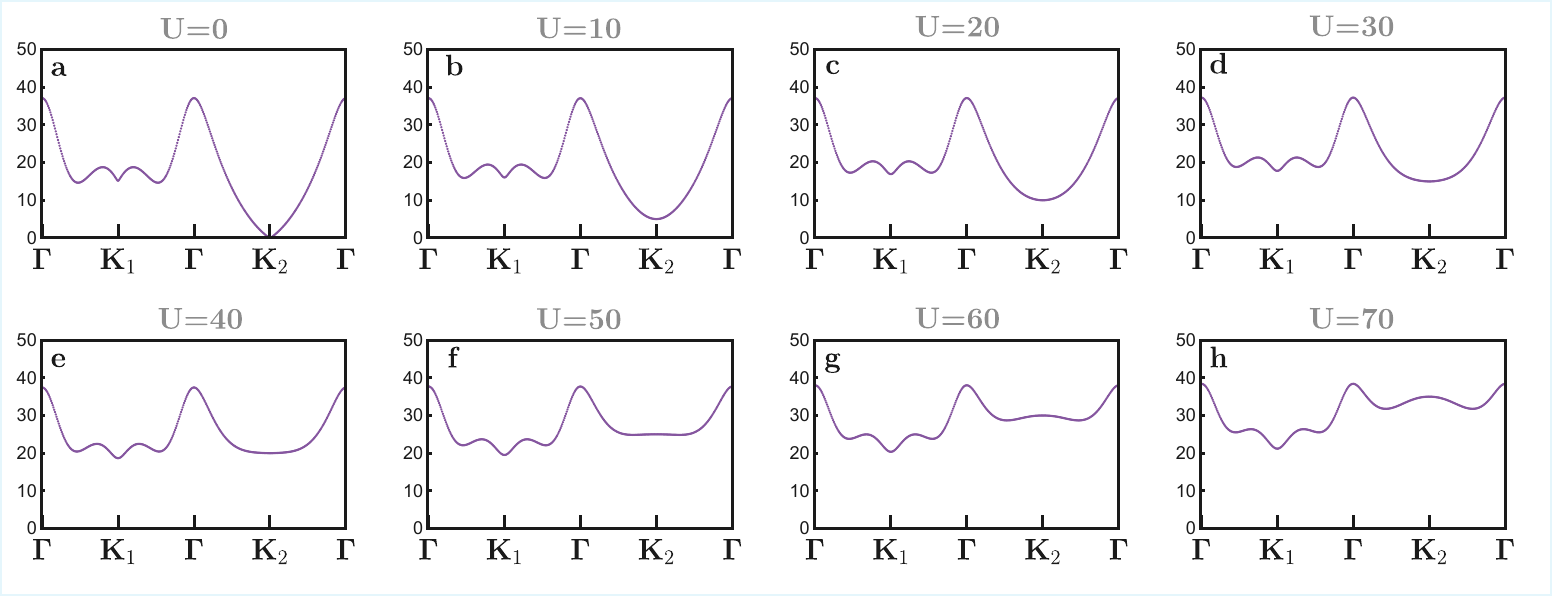}
   \caption{\textbf{Evolution of the electron band with interlayer potential difference.} 
   The lowest electron miniband is shown for increasing values of interlayer potential difference $U$ (in meV), ranging from $U=0$ to $U=70$. As $U$ increases, the bandwidth decreases and the flat portion of the band expands, reaching a maximum around $(40<U<50)$ meV, after which the band begins to buckle. Energy spectra are shown in units of meV. 
}
\label{bandsEvol}
\end{figure*}

\begin{figure*}[!t]
\centering
\includegraphics[width=1.0\textwidth]{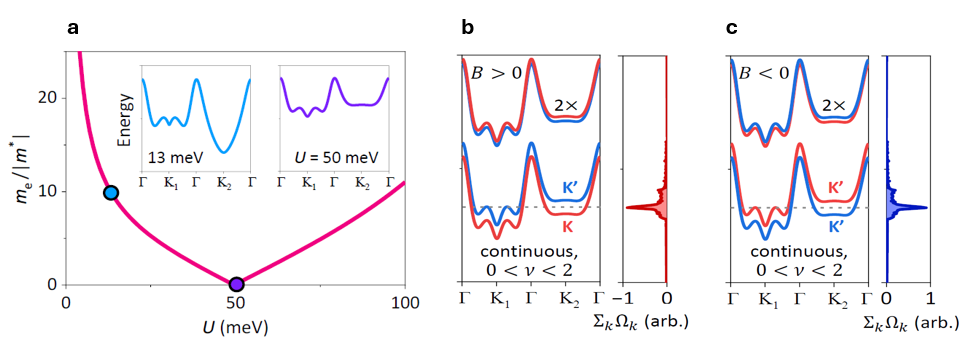}
\caption{
\textbf{Band flattening and net Berry curvature at continuous correlated states.}
(a) Inverse effective mass at the bilayer valley $K_2$ as a function of the interlayer potential energy difference $U$, induced by the applied displacement field. Insets illustrate the conduction band evolving from a dispersive shape at low $U$ to an ultra-flat portion at intermediate $U$, where the effective mass diverges. (b,c) Band diagrams at non-integer fillings ($0 < \nu < 2$) when the chemical potential resides at the locally flat band near $K_2$, for $B > 0$ and $B < 0$, illustrating continuous states. Partial band splitting occurs without a global gap. The ordering of $K$- and $K'$-derived bands reverses between opposite field directions. For $2 < \nu < 4$, the gap opens in the upper bands with reversed valley order ($K'$, $K$) for $B > 0$. The right panels show the integrated Berry curvature $\Sigma_k \Omega_k$, which peaks when the flat band is reached. The reversed $x$-axes reflect the opposite signs of Berry curvature in the two valleys and account for the antisymmetric $R_{xy}$ shifts observed in Fig.~2c,d. Red and blue denote $K$ and $K'$ valleys, with degeneracies slightly offset for clarity.
    }
    \label{valley_polarization}
\end{figure*}


\noindent \textbf{Acknowledgements}\\
We acknowledge support from the Ministry of Education, Singapore (Research Centre of Excellence award to the Institute for Functional Intelligent Materials, I-FIM, project No. EDUNC-33-18-279-V12), the National Research Foundation, Singapore, under its AI Singapore Programme (AISG Award No: AISG3-RP-2022-028), the Simons Foundation award No.~896626, the National Research Foundation of Korea (NRF) grant funded by the Korean government (MSIT) (No. RS-2022-NR071693, No. RS-2023-00303081, No. RS-2024-00444725, No. RS-2024-00410027).\\

\noindent \textbf{Author contributions}\\
K.S.N. supervised the project. M.M.E.A. and M.K. carried out the project and analysed the experimental data with the help of N.X., Y.S., S.G.X., J.B., A.B., V.F., A.K.G., S.A., and K.S.N.. S.G.X., N.X. fabricated devices. M.K., Y.S., J.B., A.B., and S.B. performed electrical measurements. M.M.E.A. and S.A. developed the theory and performed the calculations with the support of A.K., G.V., and V.F.. K.W. and T.T. supplied hBN crystals. M.M.E.A., N.X., Y.S., S.A., K.S.N., M.K. wrote the manuscript. All the authors contributed to discussions.
\\

\noindent \textbf{Data availability}\\
The data used to generate the plots in this paper can be obtained from the corresponding author upon reasonable request. \\

\noindent \textbf{Competing interests}\\
The authors declare no competing interests.\\

\noindent  \textbf{Correspondence and requests for materials}\\
\noindent should be addressed to MMEA, SA, KSN  and MK.

\end{document}